\documentclass[graybox]{svmult}

\usepackage{mathptmx}       
\usepackage{helvet}        
\usepackage{courier}       
\usepackage{type1cm}      
\usepackage{makeidx}       
\usepackage{graphicx}                      
\usepackage{multicol}       
\usepackage[bottom]{footmisc}
\usepackage{hyperref}

\makeindex            

\begin{document}

\title*{The Electron in the Maze} 

\author{Simon Ayrinhac}

\institute{Simon Ayrinhac \at Sorbonne Universit\'e, Mus\'eum d'Histoire Naturelle, UMR CNRS 7590, IRD, 
Institut de Min\'eralogie, de Physique des Mat\'eriaux et de Cosmochimie, IMPMC, 75005 Paris, France, \email{simon.ayrinhac@upmc.fr}}

\maketitle

\abstract*{This chapter presents a physical method to solve a maze using an electric circuit. 
The temperature increase due to Joule heating is observed with a thermal camera and the correct path is instantaneously enlightened. 
Various mazes are simulated with Kirchhoff's circuit laws. 
Finally, the physical mechanisms explaining 
how the electric current chooses the correct path are discussed.}

\abstract{This chapter presents a physical method to solve a maze using an electric circuit. 
The temperature increase due to Joule heating is observed with a thermal camera and the correct path is instantaneously enlightened. 
Various mazes are simulated with Kirchhoff's circuit laws. 
Finally, the physical mechanisms explaining 
how the electric current chooses the correct path are discussed.}

\section{Resolving mazes with electricity}

\subsection{Preliminary considerations about electrical circuits and thermography}

The maze-solving problem and the shortest path problem are inspiring problems in algorithmics 
and they involve many fields of science, 
such as robotics or optimization. 
In addition to numerical methods, many experimental methods have been proposed 
to solve these problems, including fluids~\cite{Fuerstman2003}, 
memristors~\cite{Pershin2011}, 
living organisms (ants~\cite{Stratton1973}, honey bees~\cite{Zhang2000}, amoeba or ``blobs"~\cite{Nakagaki2000}, nematodes~\cite{Qin2007}, plants~\cite{Adamatzky2014}) 
or plasma~\cite{Reyes2002}. 
In this chapter, a solution by a simple physical method using an electric current is proposed.

\begin{figure}[h]
  \includegraphics[width=\textwidth]{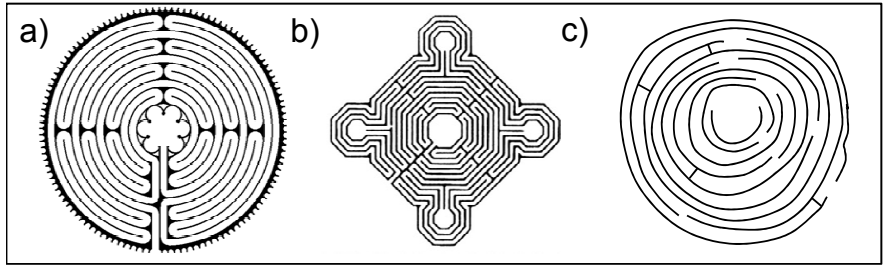}
  \caption{\label{fig:labys}Examples of labyrinths : 
	a) The labyrinth on the floor of the cathedral at Chartres (France); 
	b) The logo of \textit{Monuments historiques} (national heritage sites) in France; 
	c) A handwritten labyrinth that was designed according to the intriguing instructions ``You have two minutes to design a maze that takes one minute to solve", reproduced from the \textit{Inception} movie (real. C. Nolan, 2010).}
\end{figure}

\begin{figure}[h]
  \includegraphics[width=\textwidth]{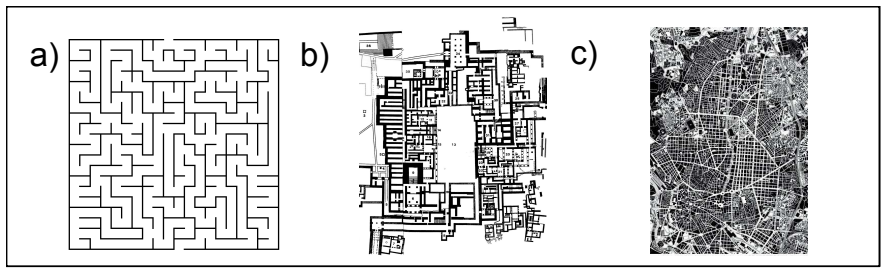}
  \caption{\label{fig:mazes}Examples of mazes: 
	a) A computer generated maze; 
	b) A plan of the Palace of Knossos (now a ruin near the town of Heraklion in Crete), the historical location of the myth of the Minotaur; 
	c) Map of a city with ways (streets and avenues).}
\end{figure}

First, mazes and labyrinths should be distinguished. 
Labyrinths have only one way, which is very complicated and which generally leads to the center, 
as can be seen in drawings on the floor of several cathedrals (see Fig.~\ref{fig:labys}). 
In contrast, mazes possess a complex branching (see Fig.~\ref{fig:mazes}). 
Although labyrinths are fascinating from a symbolic point of view, mazes are more interesting 
\footnote{The title of this chapter is a tribute to the American sci-fi writer Robert Silverberg and his novel ``The Man in the Maze".}.

This chapter presents a simple physical method to solve a maze, using an electric current. 
The maze can be done by an electrical circuit that is constituted by, for example,  copper tracks printed on an epoxy card~\cite{Ayrinhac2014}. 
Basically, two points of the maze are connected with a battery: 
if the entrance and the output of the maze are connected, then the electric current flows and 
the maze is solved. 
If they are unconnected, then the circuit is open and no current flows. 
A simple ohmmeter (usually a multimeter in a particular mode) gives the answer: 
if the resistance between two points is very low, then the path is continuous; 
in contrast, if the resistance is very high, then the path is broken. 
However, in this method, the exact path followed by the current is unknown. 

Thermography is a contactless and nondestructive method that can reveal the good path. 
The power $P$ dissipated by a resistor, with electrical resistance $R$, is given by Joule's law 
\begin{equation}
    \label{eq:jouleslaw}
    P=R I^{2}. 
\end{equation}
For a resistor obeying Ohm's law $U=RI$ the electrical energy provided by the battery is integrally converted into heat. 
When the electric charges flow, the temperature increase in the tracks is due to Joule heating.

The temperature increase $\Delta T$ is limited by thermal losses in the circuit. 
A first origin of thermal losses is conduction, which depends on the surrounding materials 
and the contact areas (controlled by the size of the circuit). 
A second origin is radiation produced by a hot body.
A third origin is the convective heat transfer between an object and the surrounding fluid---in this case, the atmosphere. 
Given that the radiation heat transfer is negligible at low temperature, 
the following simple scaling law is relevant for a standard circuit on printed circuit board (PCB)~\cite{Adam2004} :
\begin{equation}
\label{eq:scalinglaw}
    \Delta T \propto I^{2}.
\end{equation}
The increase in temperature is visualized by a thermal camera that
detects infrared radiation.

\begin{figure}[h]
\begin{tabular}{cc}
   \includegraphics[width=.45\linewidth]{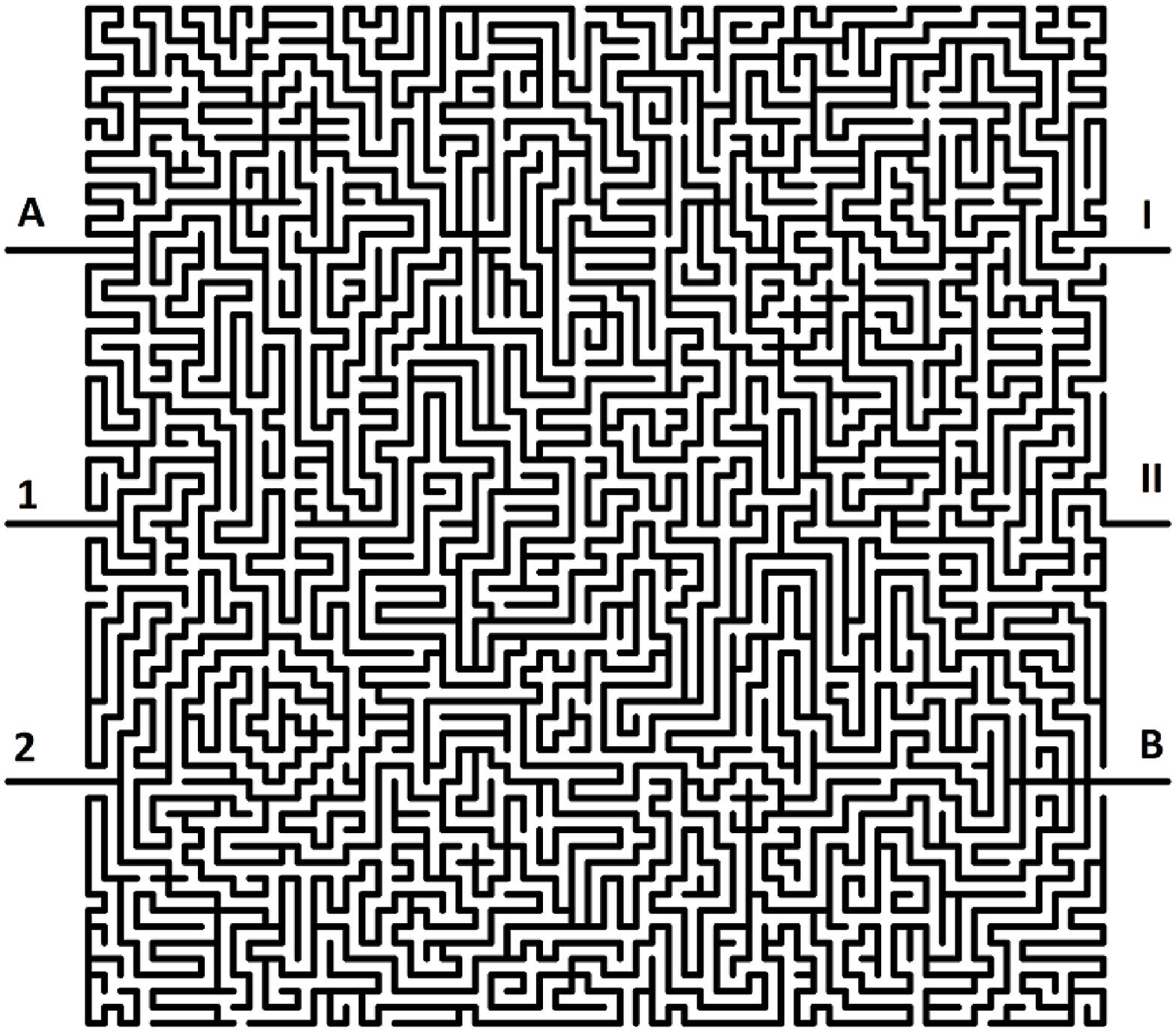} &
   \includegraphics[height=.3\linewidth]{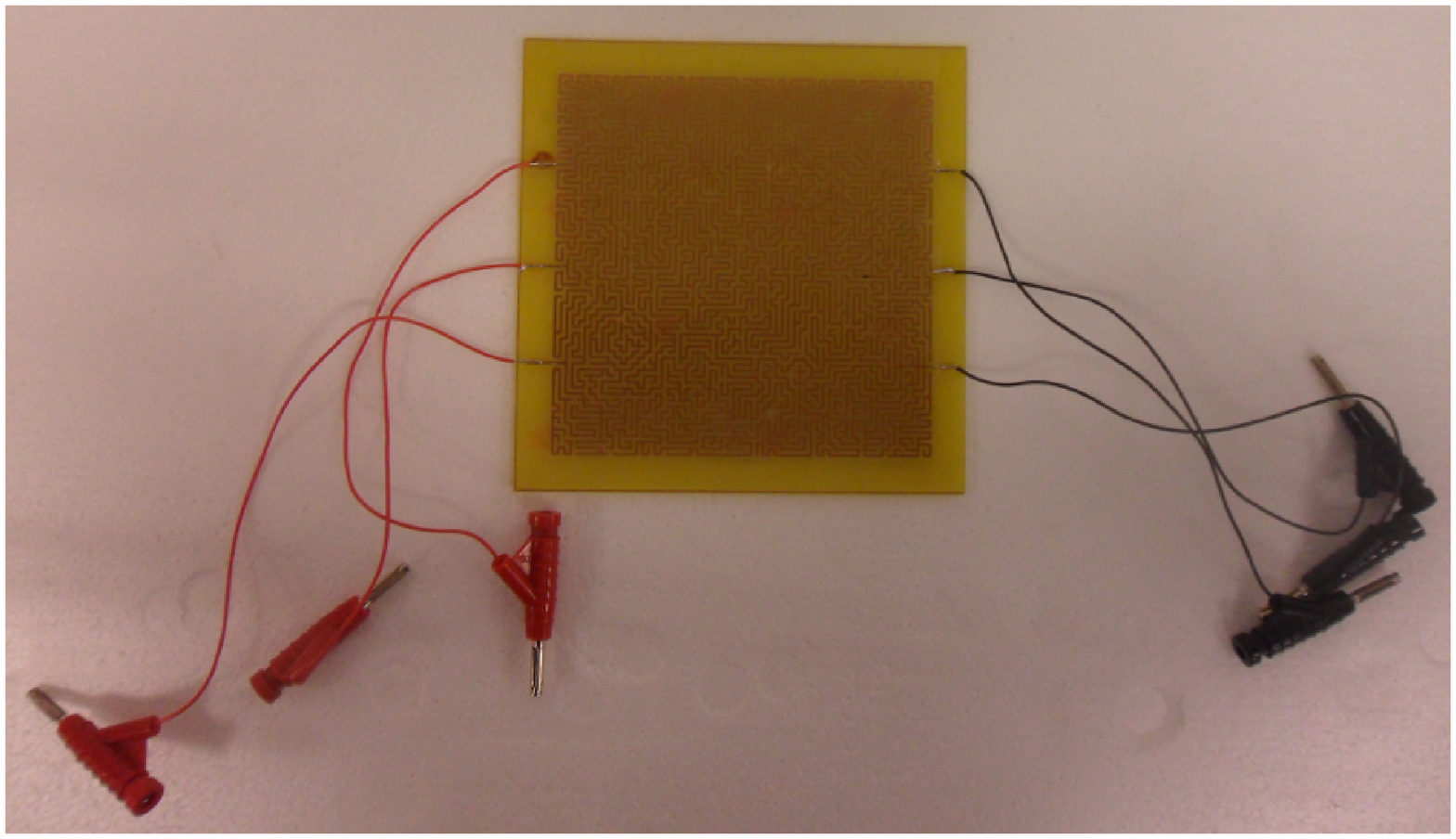} \\
\end{tabular}
  \caption{
  \label{fig:labysmall} (left) The maze used in the experiment; 
	\label{fig:circuit} (right) The maze on a printed circuit board with dimensions 15$\times$15~cm$^{2}$. 
	The conductive copper tracks have a thickness of 35~$\mu$m 
	and a width of 800~$\mu$m.
	The circuit is covered by a plastic transparent sheet.} 
\end{figure}

\begin{figure}[h]
\begin{tabular}{cc}
   \includegraphics[width=.5\linewidth]{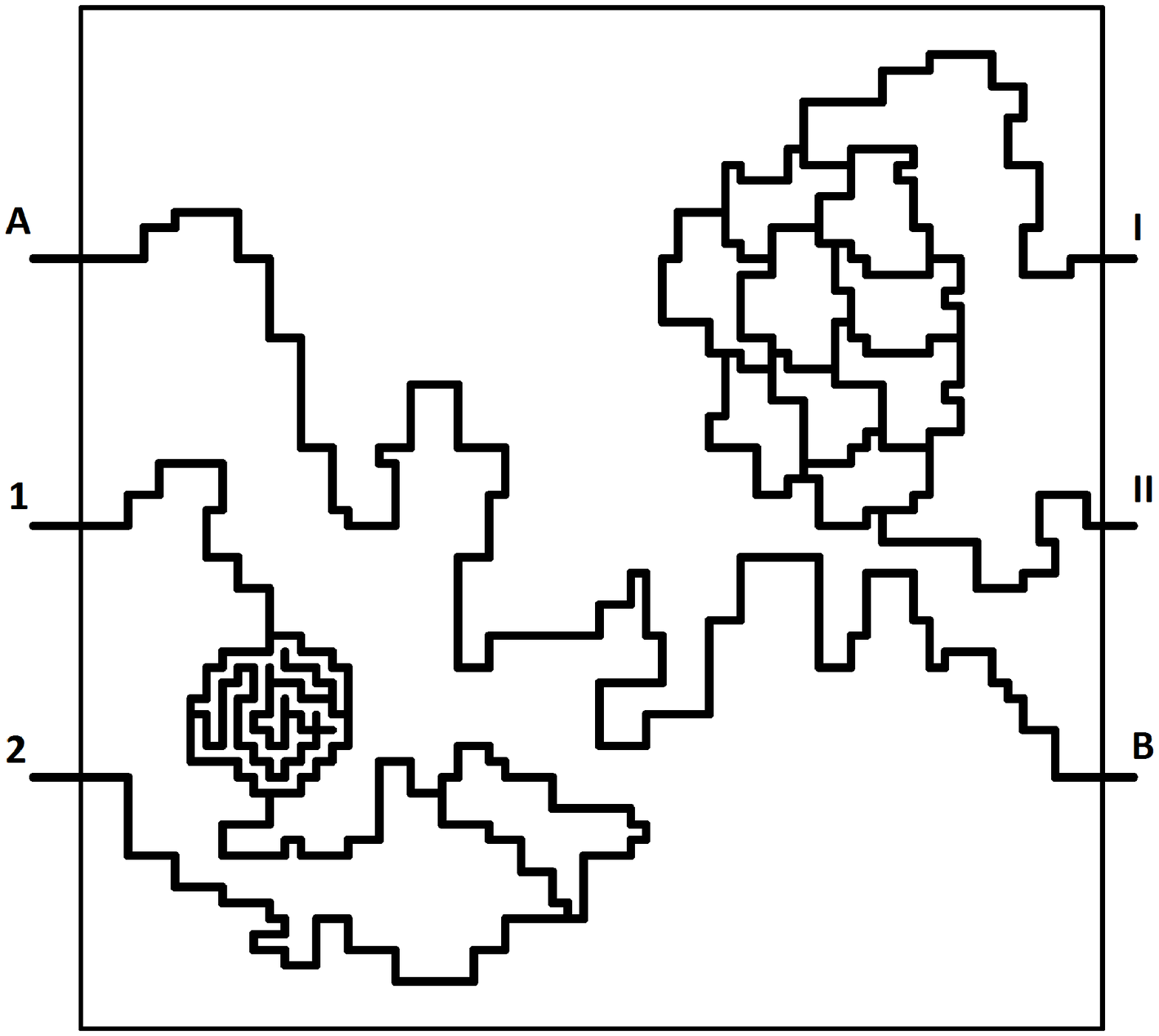} &
   \includegraphics[width=.5\linewidth]{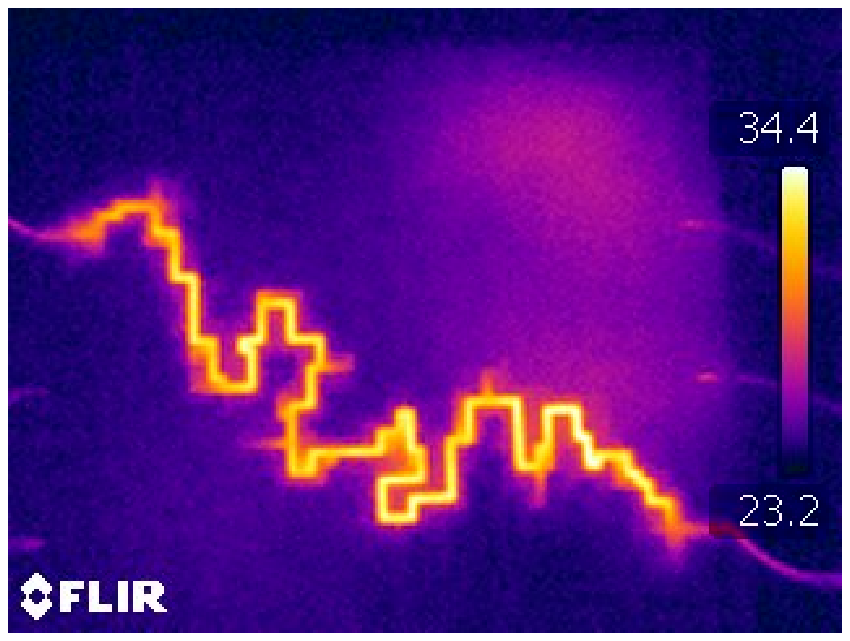} \\
\end{tabular}
  \caption{\label{fig:fig1} (left) The studied maze without dead ends ; 
	(right) Thermal image (320 $\times$ 240 pixels) of the circuit connected to the battery at points labelled (A) and (B). 
	The infrared light captured by the camera immediately shows the correct path! The colour bar on the right of the image is the temperature scale in degrees Celsius. 
	This scale is calculated with an emissivity parameter of $0.95$. 
	The spectral range of the camera is in the long-wave (LW) region (i.e. 7.5-13~$\mu$m). 
	Note that the minimum temperature on the scale is not the room temperature.}
\end{figure}

Thermal cameras are often used for educational purposes~\cite{Vollmer2001, Mollman2007, Xie2011, Haglund2016, Netzell2017}
to provide a clear visualization of invisible phenomena
or to illustrate complex phenomena. 
There a wide range of topics in physics~\cite{Vollmerbook} or in chemistry~\cite{Xie2011a} 
where a thermal camera may come in handy. 
With this kind of apparatus, qualitative as well as quantitative applications are possible. 
Although prices have decreased signiﬁcantly in recent years, thermal cameras are still rather expensive.  
However, there are other devices suitable for thermal imaging applications: such as a
simple webcam with an IR filter~\cite{Gross2005}  
or a smartphone-based device such as FLIR ONE or Seek Thermal.  

The main purpose of an infrared camera is to convert IR radiation intensity into a temperature measurement and to show the spatial variations in a false color visual image. 
Intensity is integrated from a spectral band, generally in the long-wave infrared (7.5-13 µm) region. 
The temperature is given by a formula accounting for three phenomena~\cite{Vollmerbook}: the true thermal emissions from the object, 
the thermal radiation emitted by its surroundings and reﬂected by the object, and the atmospheric absorption. 
For a proper temperature measurement,  knowledge of certain parameters (e.g., emissivity, humidity, distance, and ambient temperature) is necessary.

Thermal imaging can find a wide application in electronics. For example, electrical components in the microelectronic boards of computers produce heat that can damage the circuits. To avoid failures, processors or power transistors need to be cooled by fans or Peltier modules, for example. IR imaging is a non-contact and non-destructive technique that can be used to test and survey electronic boards, allowing a diagnostics of possible malfunctions. Given that these boards are often made of different materials, the differences in components emissivity makes quantitative temperature measurements difficult (an explanation of emissivity will be given later). 

Temperature measurement depends strongly on the emissivity $\epsilon$ of materials. 
Unfortunately, for metals, the emissivity is very low, and they are hard to see directly in thermography. 
Emissivity is defined as the ratio of the amount of the radiation emitted 
from the surface to that emitted by a blackbody at the same temperature~\cite{Vollmerbook}. 
A blackbody is a perfect absorber for all incident radiation. It appears black when cooled at 0~K and 
when heated up it emits light at all wavelengths and the resulting spectrum 
(given by Planck's law~\cite{Vollmerbook}) depends only on the temperature of the blackbody. 

Kirchhoff's law of thermal radiation states that $\epsilon=\alpha$ 
where $\alpha$ is the absorption coefficient~\cite{Besson2009}. 
This formula means that, for an opaque body, the more a body absorbs, the more it emits light. 
A metal is a good reflector, so it has a bad absorption and, therefore, a poor emissivity.

\subsection{Experiments on circuits}

A regular maze (see Fig.~\ref{fig:circuit}) is printed on a epoxy card with tracks made of copper. 
A transparent plastic sheet, which possess a higher emissivity than bare metal, 
is placed on top of the circuit to ensure that the temperature increase is seen by thermography. 
The transparent cover sheet allows the maze to be seen in both in infrared light and visible light. 
With a thermal camera, the correct track appears to be immediately illuminated, 
despite the complexity of the circuit (see Fig.~\ref{fig:fig1}).

\begin{figure}[h]
\begin{tabular}{cc}
   \includegraphics[width=.45\linewidth]{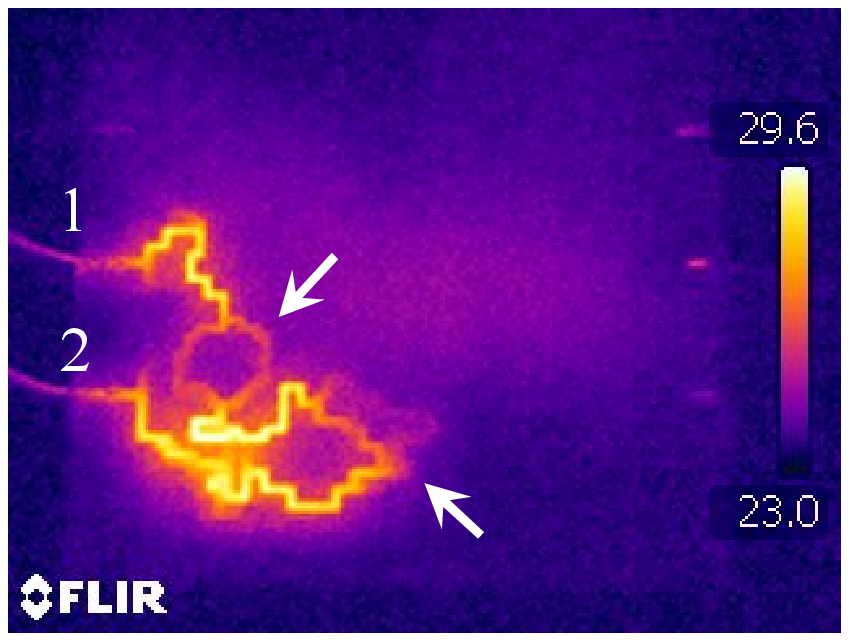} &
   \includegraphics[width=.45\linewidth]{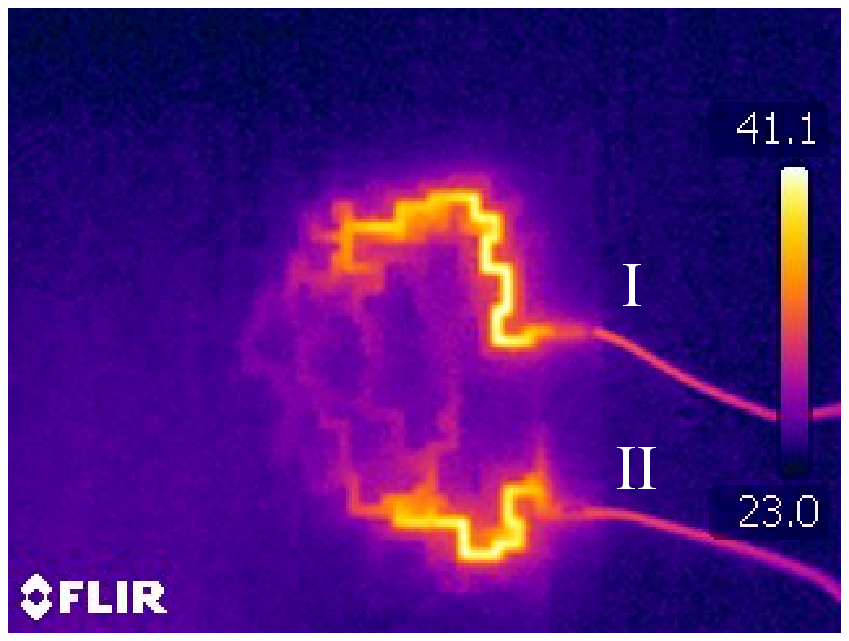} \\
\end{tabular}
  \caption{\label{fig:figs23}
  (left) The battery is connected to the circuit by the points labelled (1) and (2). 
	This picture illustrates the consequences of two particular circuit configurations: 
	the difference between two paths with a path with double length compared to another (lower arrow) 
	and the Wheatstone Bridge (upper arrow).  
	The battery (right)  is connected to the circuit by the points labelled (I) and (II). 
	Because there are many good branches with equal lengths, it is difficult to identify the shortest path.}
\end{figure}

Our maze is designed to highlight the following special features (see Fig.~\ref{fig:figs23}): 
\begin{itemize}
  \item In case of branching with a path twice as long as the parallel branch, 
the shorter path appears to be more brightly illuminated compared to the longer path. 
This happens because the trace 
resistance $R$ is proportional to the length of the resistor $\ell$, such as 
$R = \ell / \sigma A$, where $\sigma$ is the electrical conductivity and $A$ the cross-sectional area.  
The voltage $U$ is equal in the two branches and gives $I_{\ell}=2I_{2\ell}$,
so with equation~(\ref{eq:scalinglaw}), 
the temperature increase in the shortest path is four times higher than in the longer path $\Delta T_{\ell}=4\Delta T_{2\ell}$. 
  \item If the branching is configured as a Wheatstone bridge, then
the parallel branch does not appear. 
See Fig.~\ref{fig:simu9} and the associated text for explanations. 
  \item In case of multiple paths, the branching is complicated and the shortest path is hard to be seen. 
\end{itemize}

This kind of demonstration is possible provided that several conditions are met: the tracks should have the same section and they should be built with the same material, the branching should not be too complex (i.e., one-solution mazes) and the correct path should exist among many dead ends. So, the ideal circuit is an intermediate between a labyrinth and a maze.

\begin{figure}
\begin{tabular}{cc}
   \includegraphics[width=.4\linewidth]{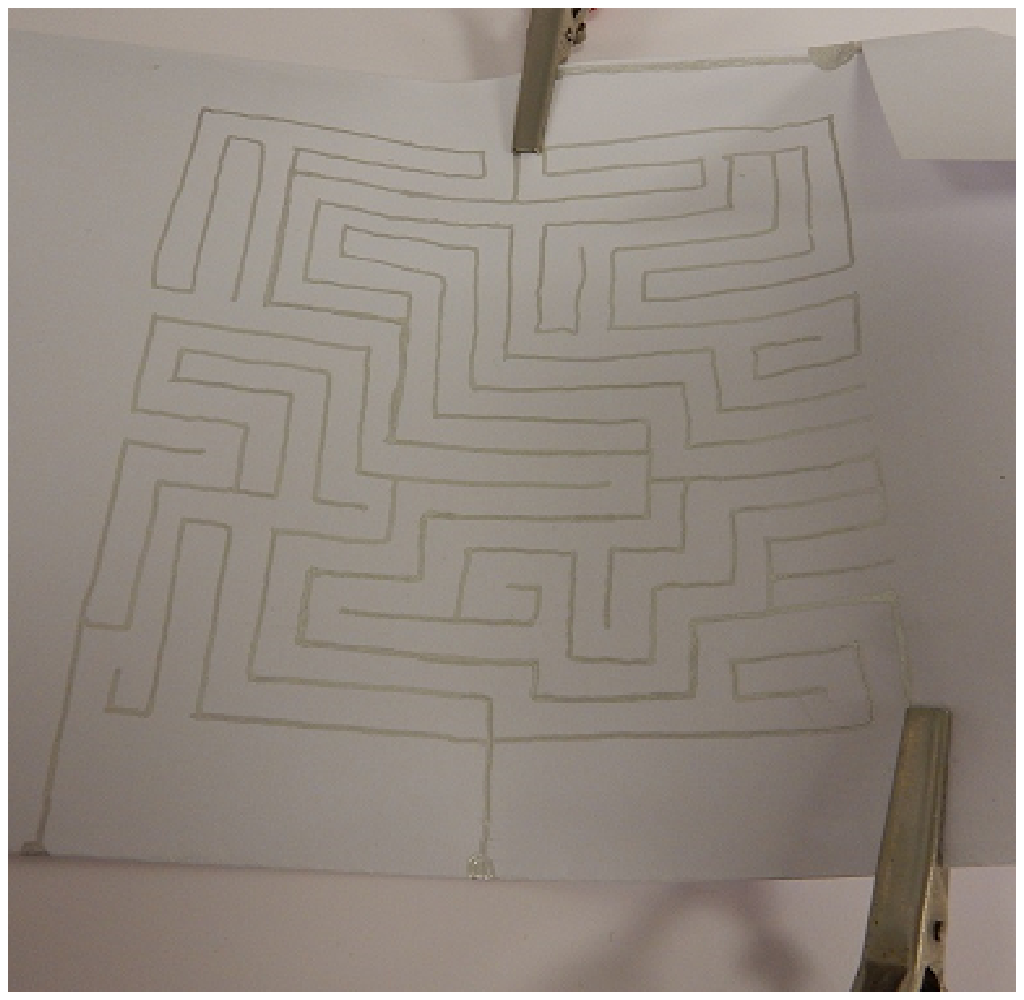} &
   \includegraphics[width=.5\linewidth]{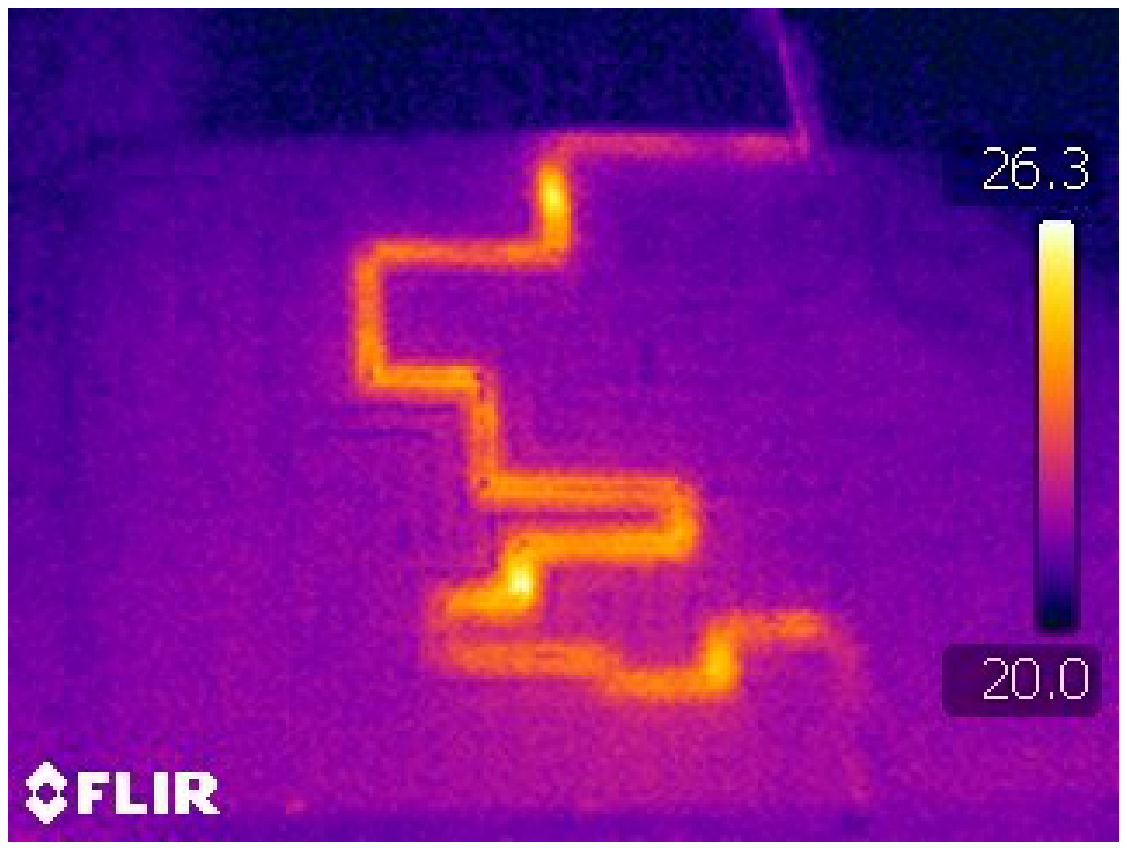} \\
\end{tabular}
  \caption{\label{fig:papermaze}
  (left) A maze is drawn on a paper sheet with a pen delivering conductive ink (around 1~$\Omega$/cm).  
	 It takes about 10 minutes to draw the whole maze. 
	This maze was previously presented in detail in Ref.~\cite{Ayrinhac2014}. 
	Despite the care taken in the drawing, the tracks are not perfectly regular 
	and the paper is much more fragile compared to a PCB, and can be torn easily 
	(right). The correct path appears illuminated with an infrared camera. 
	The temperature increase is clearly seen in this case because the paper has a higher emissivity compared to metallic conductive tracks.
	Due to the sideways spreading of the heat, the correct path in the image looks ``blurred". 
	This method is cheaper and faster compared to printing the same maze on a PCB.}
\end{figure}

\begin{figure}
\begin{tabular}{cc}
   \includegraphics[width=.4\linewidth]{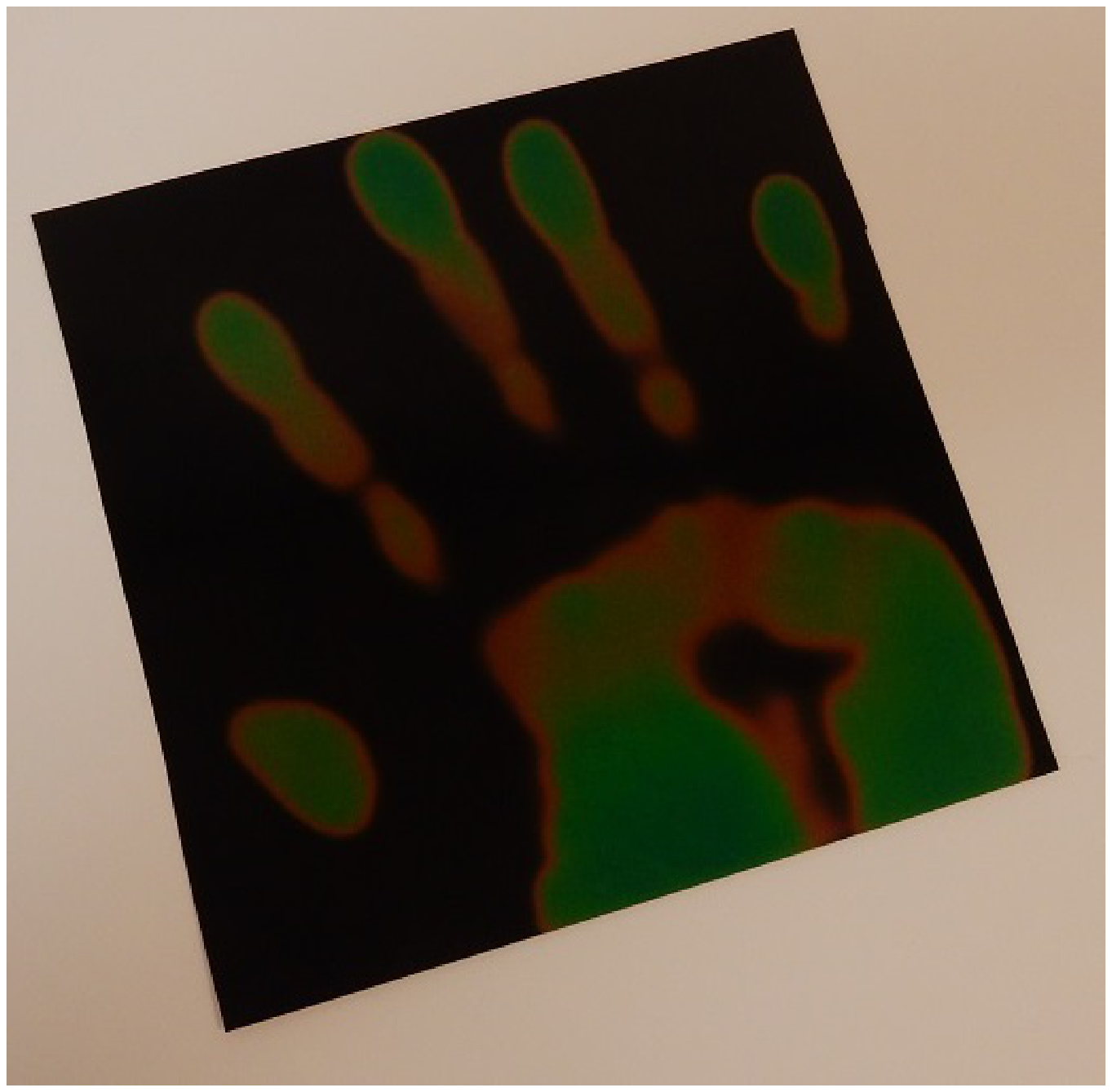} &
   \includegraphics[width=.5\linewidth]{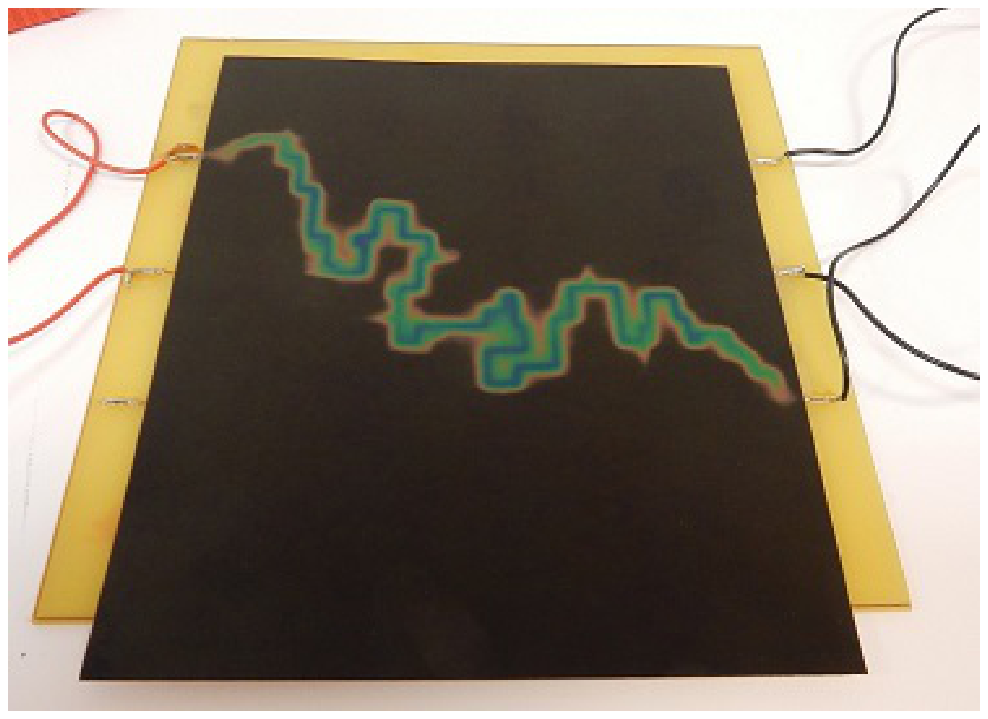} \\
\end{tabular}
  \caption{\label{fig:thermalsheet}
  (left) A thermochromic liquid crystal film (the size of the sheet is 15$\times$15 cm$^{2}$), sensitive to temperature, 
	with a hand print. 
	The transition from black to color occurs between 20--25$^{\circ}$C. 
	The colour change is reversible and quick, with a response time about 10~ms~\cite{Stasiek2014}. 
	(right) The correct path of the maze appears illuminated with the liquid crystal film placed on the PCB.}
\end{figure}

The circuit can be drawn by an ink pen on a paper sheet (see Fig.~\ref{fig:papermaze}). 
However, despite the care taken in the drawing, the tracks are not perfectly regular and the paper is much more fragile than a PCB and can be torn easily. 
Nevertheless, this method is cheaper and faster than printing the same maze on a PCB. 
To further reduce costs, the IR camera can be replaced by a temperature sensitive liquid crystal film (around 15\$), 
to obtain qualitatively the same result (see Fig.~\ref{fig:thermalsheet}).  

\subsection{Simulated circuits}

To investigate more complex topologies, various circuits were simulated 
in the permanent regime using Kirchhoff's laws~\cite{Chabaybook} : 
the algebraic sum of currents at a node is zero (Kirchhoff node rule), 
and the directed sum of the voltages around a loop is zero (Kirchhoff loop rule). 
The operation involves a solution of a linear system  
involving resistances, currents and applied voltages 
with $n$ equations, where $n$ is the number of branches in the electrical network.

\begin{figure}[h]
\begin{tabular}{ccc}
   \includegraphics[width=.2\linewidth]{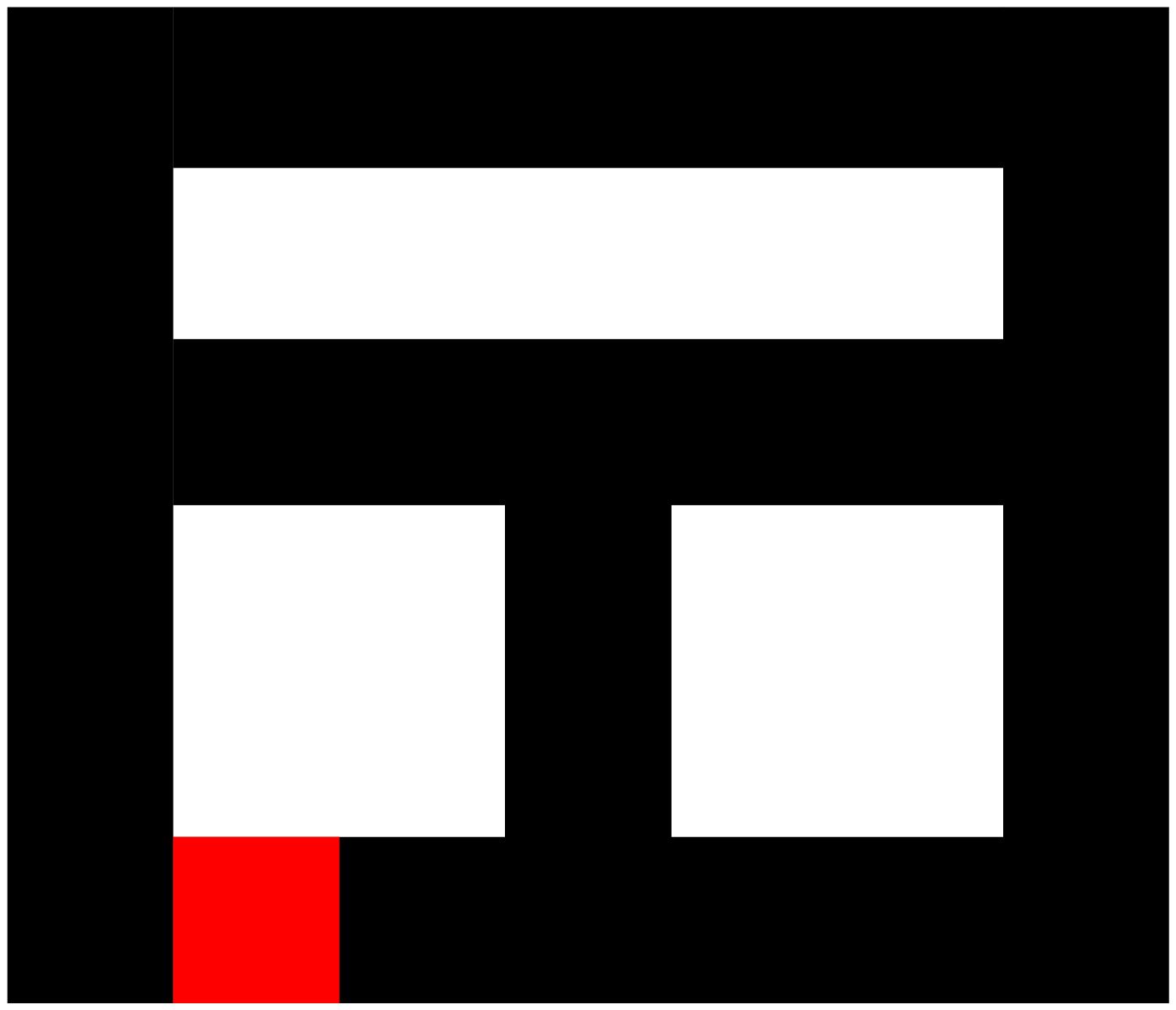} ~~~~~~~~~~&
   \includegraphics[width=.3\linewidth]{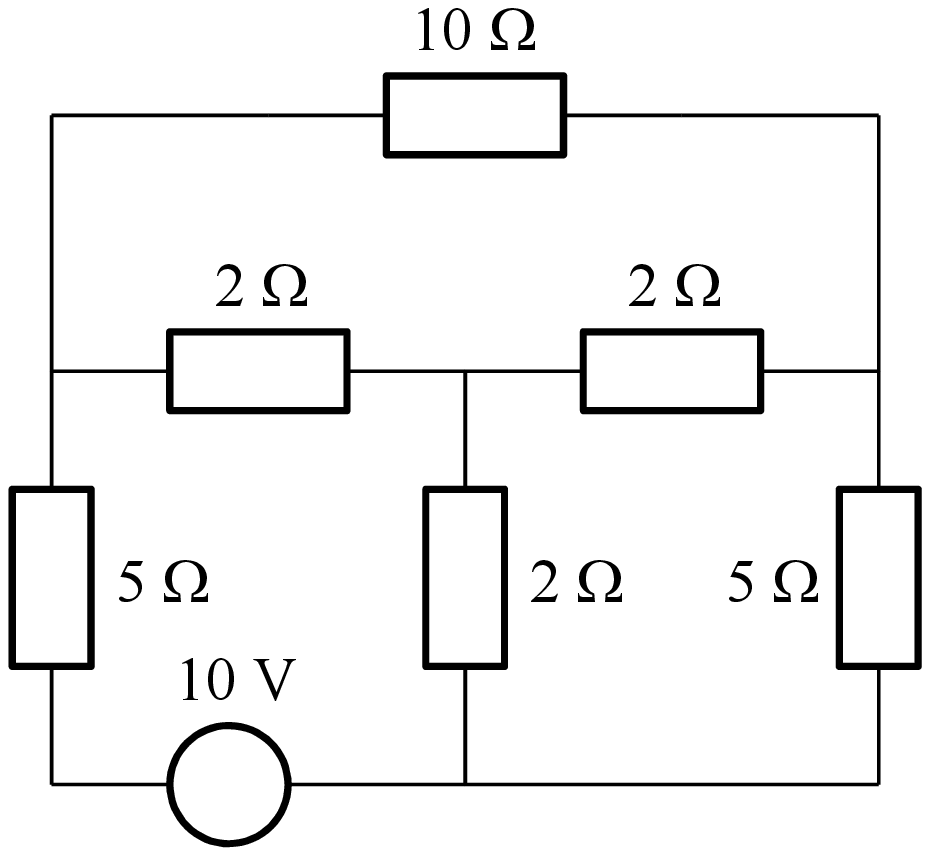} ~~~~~~~~~~&
	 \includegraphics[width=.3\linewidth]{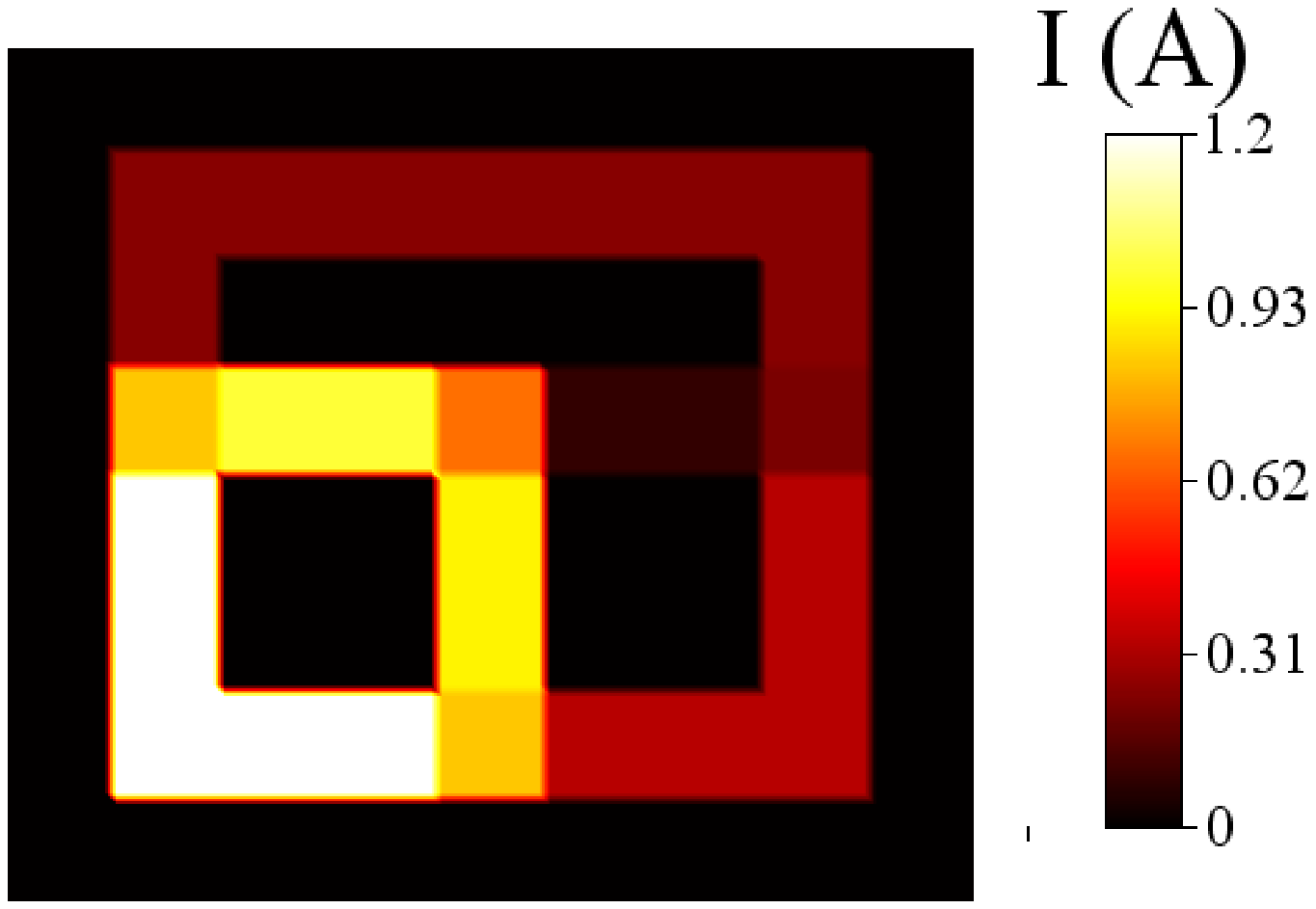} \\ 
\end{tabular}
  \caption{\label{fig:circuit_equiv}
  (left) A circuit drawn in an image file (9$\times$8 pixels), imitating copper tracks on a PCB. 
	(center) The equivalent electrical circuit. 
	(right) Distribution of the intensities obtained by the application of Kirchhoff's laws (see text). 
	The color bar at the right indicates the intensity values in each branch. }
\end{figure}

The studied circuits are directly generated by drawing the tracks in an image file 
(see the example in Fig.~\ref{fig:circuit_equiv}). 
The battery voltage is 10~V and the track electrical resistance is 1~$\Omega$ by unit of length 
(equal to the track width). This resistance value is arbitrary. 
The nodes are not be taken into account in the calculation of resistance.  
In the nodes, the current is calculated by the averaging of the surroundings currents. 
The resulting picture represents the current $I$ in amperes at each point of the circuit. 
Note that the resulting picture is not a thermography image rendering. 

\begin{figure}[h]
\begin{tabular}{cc}
   \includegraphics[width=.3\linewidth]{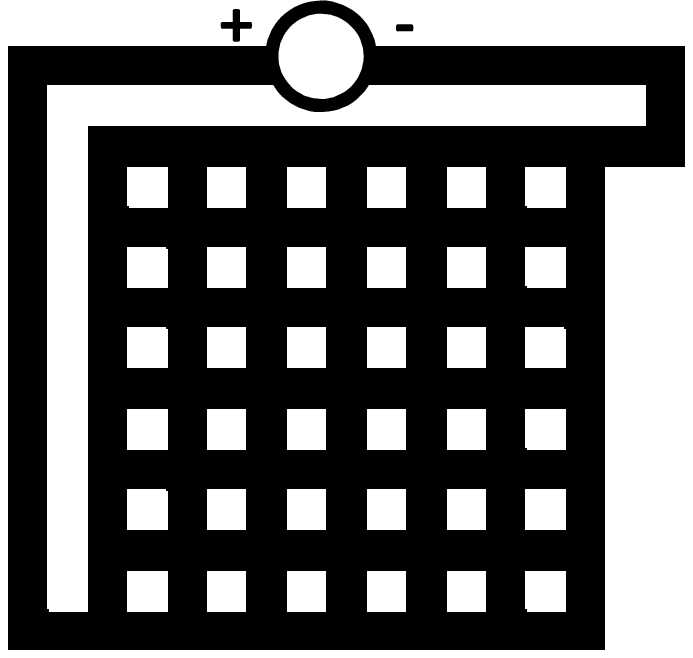} ~~~~~~~~~~~~~~~~&
   \includegraphics[width=.4\linewidth]{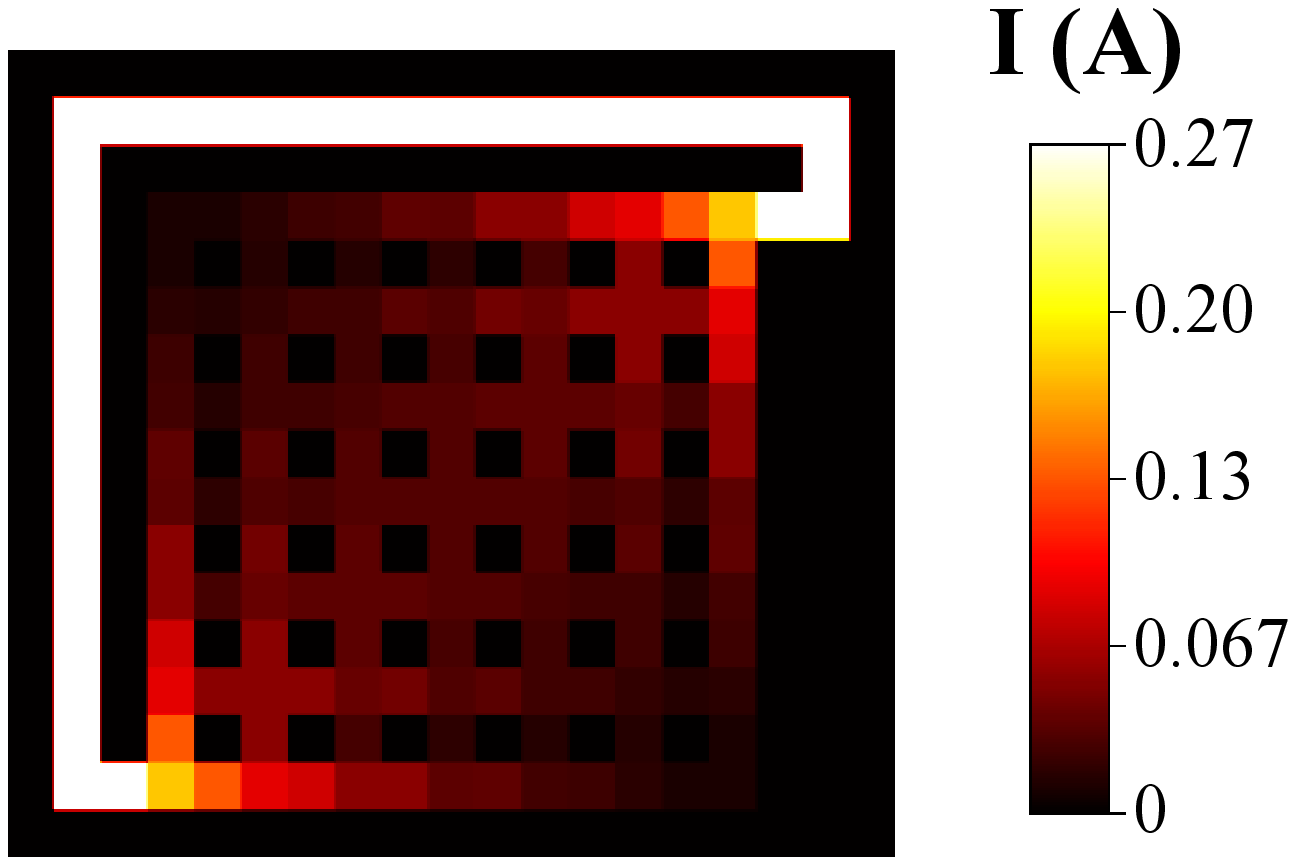} \\
\end{tabular}
  \caption{\label{fig:simu18}
  (left) A grid-like circuit. The battery is located in the upper branch 
	(right). Distribution of the intensities inside the studied circuit.  
	The color bar at the right indicates intensity values in each branch.}
\end{figure}

\begin{figure}[h]
\begin{tabular}{cc}
   \includegraphics[width=.3\linewidth]{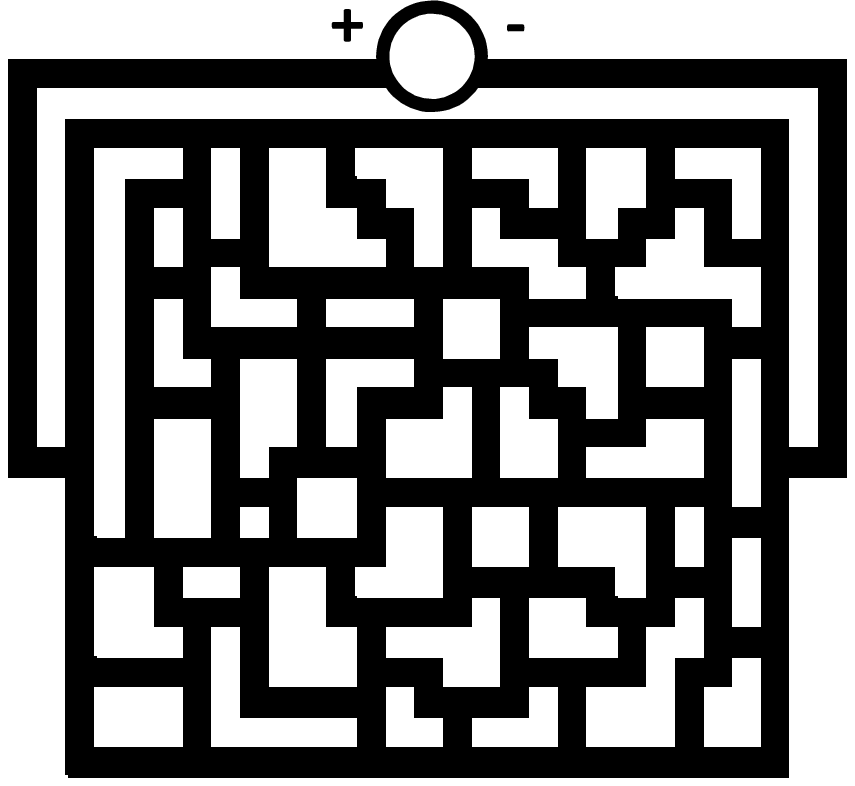} ~~~~~~~~~~~~~~~~&
   \includegraphics[width=.4\linewidth]{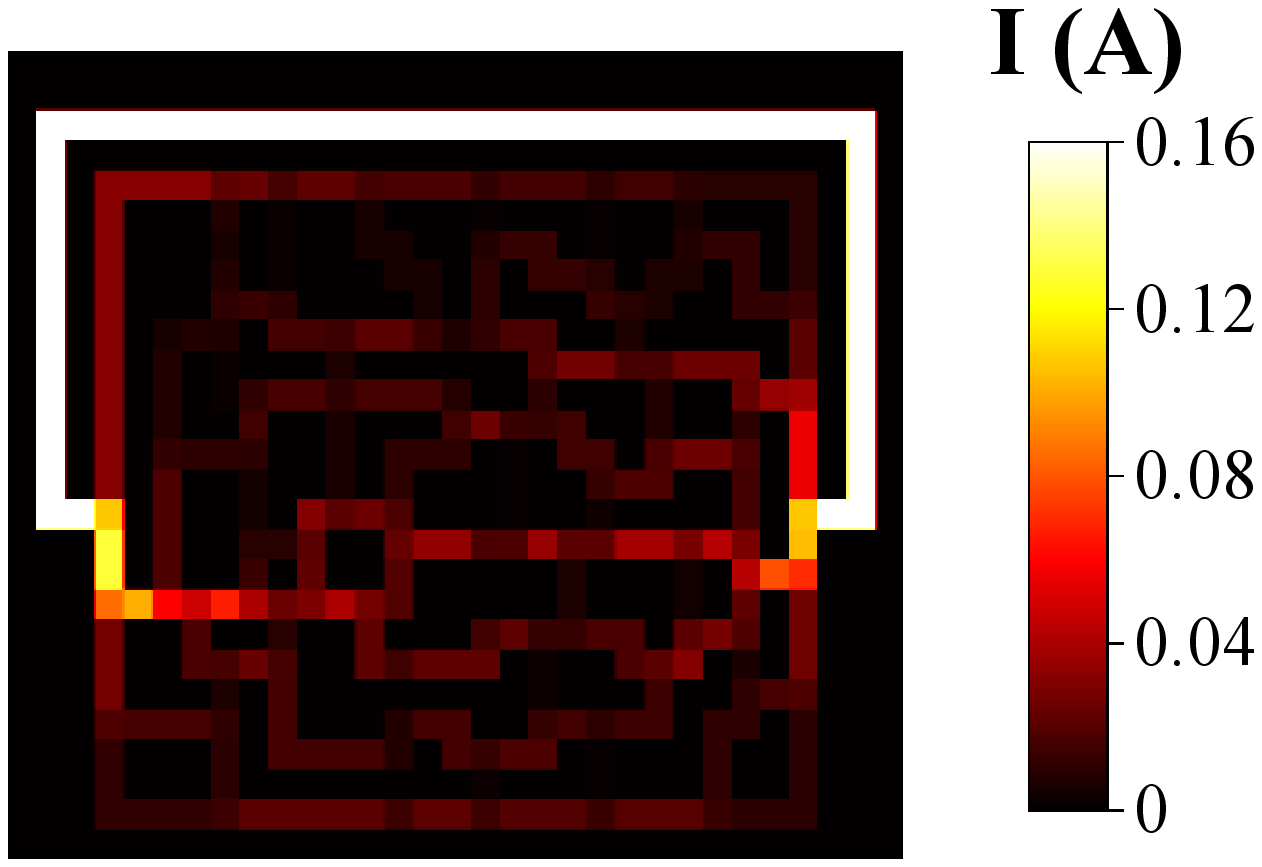} \\
\end{tabular}
  \caption{\label{fig:simu13}
  (left) A circuit maze with disordered tracks. The battery is located in the upper branch 
	(right). Distribution of the intensities inside the studied circuit. The color bar at the right indicates the intensity values. }
\end{figure}

\begin{figure}[h]
\begin{tabular}{ccc}
   \includegraphics[width=0.25\linewidth]{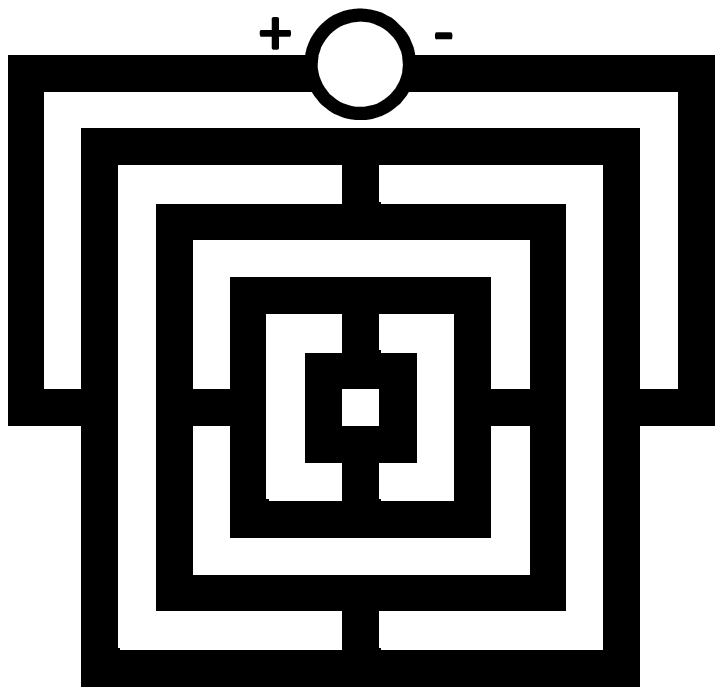} ~~~&
   \includegraphics[width=0.35\linewidth]{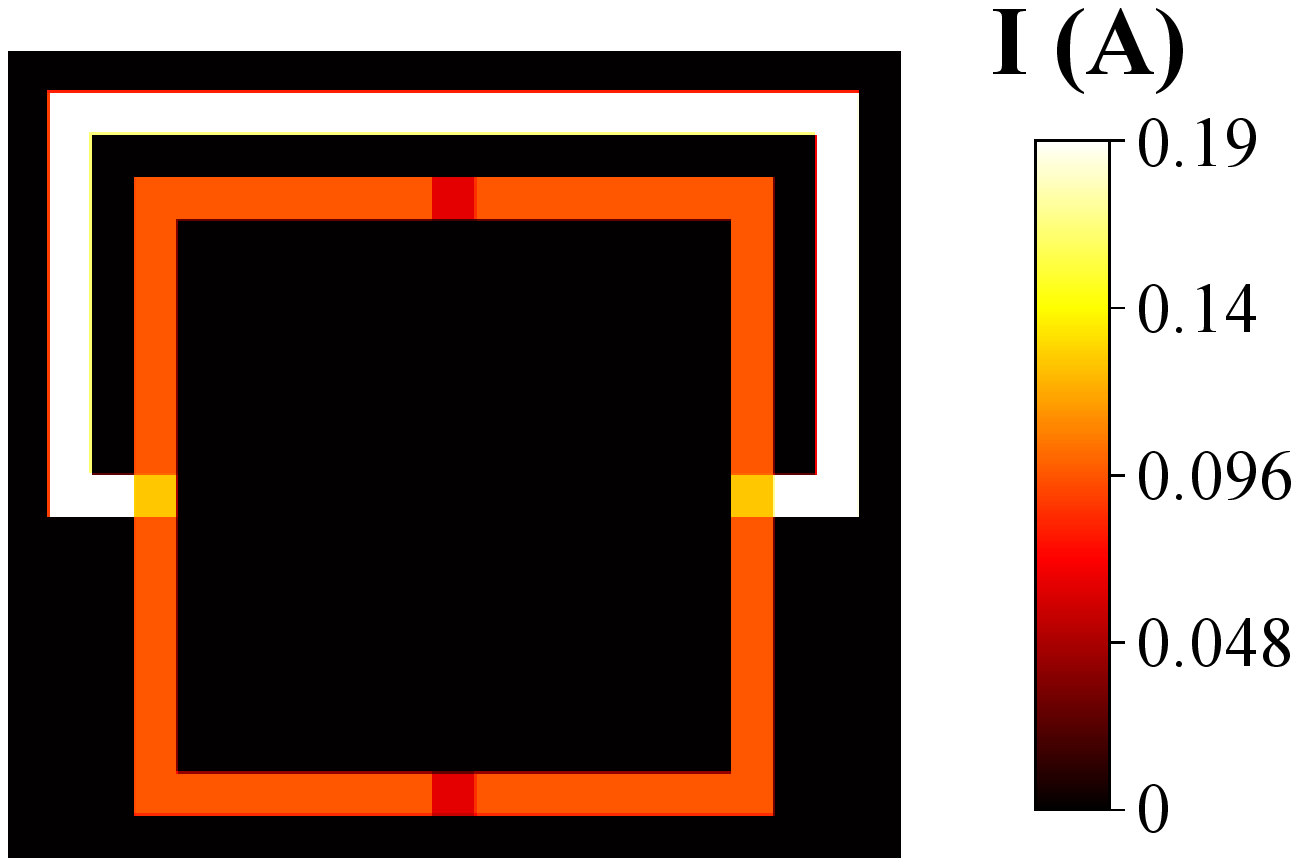} ~~~&
	 \includegraphics[width=0.3\linewidth]{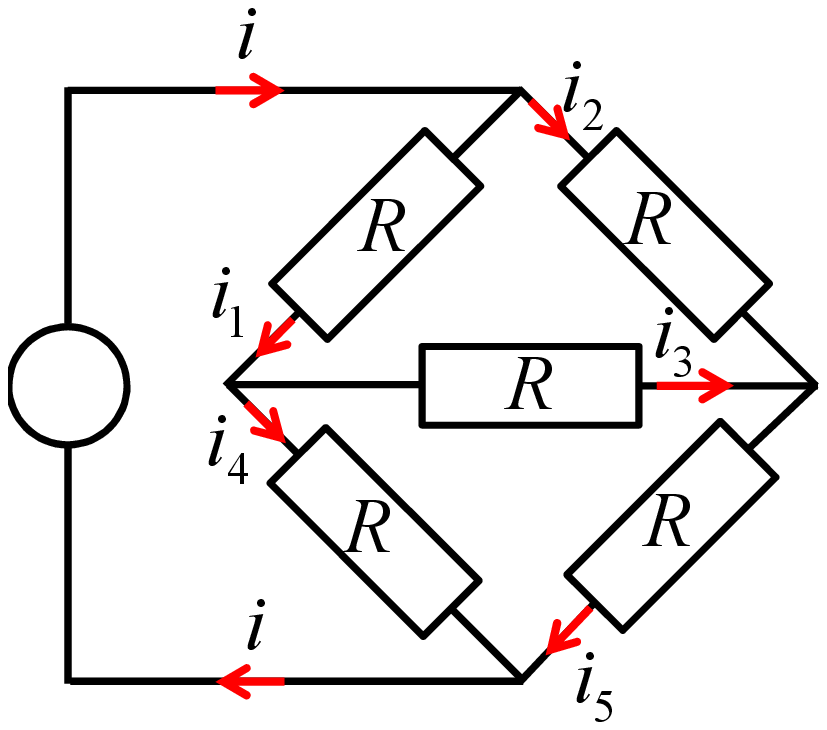}  \\
\end{tabular}
  \caption{\label{fig:simu9}
  (left) A complex circuit. The battery is located in the upper branch
	(center). In the simulated circuit, the central part does not appear.  
	This is due to the creation of an electrical ``bridge"
	(right). Diagram of an electrical circuit containing an electrical ``bridge". 
	Resistors and currents are labelled $R$ and $i_{n}$, respectively. 
	If the circuit is well-balanced, then all $R$ are equal, and it can be demonstrated that $i_{3}=0$.
	This configuration is called a Wheatstone bridge and it is often used to measure resistance.}
\end{figure}

In a grid-like circuit (Fig.~\ref{fig:simu18}), the current is spread over the whole circuit. 
In this case, all of the paths are equivalent and the current appears equal in all of the paths.  
In another example with disordered tracks (Fig.~\ref{fig:simu13}), the shortest path appears clearly. 
In some cases, the current can fall to zero in a part of the circuit. This is due to the creation of an ``electrical bridge", also called Wheatstone bridge, 
as illustrated in the Fig.~\ref{fig:simu9}. 
The process to follow the shortest path between two points connected by the battery is to choose 
at each node the branch where the intensity is maximum.
Generally speaking, the shortest path is the path where the intensity is maximized. 
This idea is sustained by the basic electric conception that more current follows the path of less resistance.

The resistive grid was early used to explore some physical problems, such as solution of partial differential equations~\cite{Liebmann1950}, 
or mobile robot path planning~\cite{Tarassenko1991}. 
In robot path planning, a collision-free environment can be modelled with a resistive grid of uniform resistance, 
and obstacles are represented by regions of infinite resistance. 
The path planning can be evaluated in real space if the robot moves through a maze, for example, 
or in the configurational space where the dimensions are the degrees of freedom of the robot, considering a robot manipulator arm, for example. 
The path from start to goal is found using voltage measurements from successive nodes. 
In the limit of the continuous case, if we assume that conductivity is uniform and constant, 
then the electromagnetism equations imply that for steady currents (in 
regions where there is no sources) the electric potential $V$ obeys Laplace's equation. 
The two-dimensional Laplace's equation 
\begin{equation}
   \frac{\partial^{2} V(x,y)}{\partial x^{2}}+\frac{\partial^{2} V(x,y)}{\partial y^{2}}=0,
\end{equation}
may be solved to calculate the electric potential $V$ at every point $(x,y)$. 
The direction of the movement is given locally by the direction of the voltage gradient $\vec{\nabla} V$.
Globally, this approach produces an optimal path solution, depending on the limit conditions to avoid spurious local minima.

\section{Discussion of the physical mechanisms}

After the experimental demonstration, a physical question remains: 
How does the electric current choose the correct path amongst many others?

A common explanation is that the battery produces a potential difference $\Delta V$ at the two extremities of the circuit and 
the resulting electric field $\vec{E}$ possesses a magnitude constant and a direction along the wire. 
This is especially puzzling in a maze circuit, where the electric field must follow the multiple bends of the circuit. 
In contrast, the electric field $\vec{E}$ and then the potential difference $\Delta V$ is produced by distributions of point charges. 
So, where are the charges producing the electric field inside the wires? 
This question is challenging because \textit{electrokinetics} and \textit{electrostatics} 
are two topics that are usually treated separately in physics textbooks: 
charges distributions on one side, and electrical circuits on the other side. 
Consequently, for most students, the two topics are unconnected, leading to many misconceptions~\cite{Rainson1994};  
that is, commonly held beliefs that have no basis in actual scientific knowledge. 

This point was extensively examined in the literature~\cite{Rosser1970, Heald1984, Jackson1996, Preyer2002, Muller2012}. 
The electric charges responsible for the electric field inside the conductor are located on the surface of the wire. 
This fact is known from the pioneering works of Weber and Kirchhoff~\cite{Assis2007}, but it has been completely forgotten during the last 150 years. 
The quantity of surface charge is very small: the order of magnitude of the charge necessary to turn an electric current 1~A around a corner 
is equal to about the charge of one electron~\cite{Rosser1970}. 
A quantitative estimate in a typical circuit~\cite{Muller2012} for the magnitude of the surface charge density 
is $10^{-12}$-$10^{-10}$~C$\cdot$m$^{-2}$. 
Comparatively, the quantity of charge moving inside the wires is much higher, about 10$^{-6}$~C$\cdot$m$^{-2}\cdot$s$^{-1}$, which corresponds to 1~A.

Although the quantity of surface charges is small, their role is essential \cite{Jackson1996}: 
they ensure that the equality of the potential in the conductors is at equilibrium,  
they permit the circulation of the charges 
and they produce an electric field outside the wires. 
Two types of surface charges can be distinguished \cite{Muller2012}: 
at the boundary of two conductors with different resistivities
and at the surface of the conductors. 

The free electrons in the metal are pushed by the electric force arising from the electric field. 
If there is a curve, then they pile-up on the surface and their electric field changes the pathway of the incoming moving charges. 
There is a \textit{feedback mechanism} between the surface charges and the charges moving inside. 
Chabay and Sherwood \cite{Chabaybook} provide an excellent and very accessible overview 
of this problem for undergraduate students.

This feedback mechanism explains how the charges avoid dead-ends of the circuit maze. 
First, they pile-up on the extremity of the dead-end;
the build-up of negative charge pushes the arriving electrons and 
then the flowing current reaches zero. 
Finally, the only path left for the charges to follow is the solution path of the maze. 
This phenomenon is analogous to liquid propagating in a microfluidic network~\cite{Fuerstman2003}.

Electric circuits are often compared to hydraulic circuits from a pedagogical point of view 
(the most complete comparison can be found in Table~1 in Ref.~\cite{Oh2012}). 
However, there are fundamental differences between electrons and water: 
electrons do not interact with one another 
and energy is not carried by the free electrons. 
Energy is carried outside the circuit by the electromagnetic fields 
forming the Poynting vector $\vec{S} = \frac{1}{\mu_{0}} \vec{E} \times \vec{B}$. 
This formula combines the magnetic field $\vec{B}$ due to electric current inside the wires (moving charges) 
and the electric field $\vec{E}$ due to surface charges.

Solving the maze with an electric current 
reveals the existence of a transient period between the beginning of the experiment 
and the moment that the current is stabilized in the solved maze  
\footnote{In the following discussion, we do not consider the thermal equilibration of the system, 
which requires more time than electric equilibration.}. 
Simulations performed by Preyer in a simple RC circuit~\cite{Preyer2002} 
can give us a better understanding of phenomena observed in the transient state.  
Just after the connection of the battery at two points of the maze,
the electric field spreads through the circuit at the speed of light. 
During this step, the surface charges  build on the tracks. 
The surface charges locally change the electric field 
and the current because they influence each other. 
This feedback mechanism occurs in the transient state and is at work when the uniform current flow is established, 
which means that the maze is solved.  

All of the changes occur at the speed of light $c$ inside the material around the circuit (usually air), 
which is also the speed that the information propagates between different parts of the circuit. 
The drift velocity $v$ of charges moving inside the wires is much slower than $c$, 
typically a few microns a second. 
The simple propagation of the electric field through the whole circuit needs a time $\tau \approx \ell / c$, 
where $\ell$ is the characteristic length of the maze, 
but the time $\tau'$ needed for the feedback mechanism to operate is much longer~\cite{Preyer2002} with $\tau' > 2 \ell / c$, 
which can be considered as the minimum time needed to solve the maze. 

With this physical method, the maze resolution is fast. This explains why 
this is considered to be the fastest and the cheapest method among many other physical methods~\cite{Adamatzky2017}, 
especially if the circuit is drawn by ink pen on a paper sheet.

\section{Acknowledgements}

The thermal camera was provided by the ``UFR de physique" at UPMC. 
The author is indebted to M. Fioc for his valuable comments on the manuscript.


\begin{thebibliography}{10}
\providecommand{\url}[1]{{#1}}
\providecommand{\urlprefix}{URL }
\expandafter\ifx\csname urlstyle\endcsname\relax
  \providecommand{\doi}[1]{DOI~\discretionary{}{}{}#1}\else
  \providecommand{\doi}{DOI~\discretionary{}{}{}\begingroup
  \urlstyle{rm}\Url}\fi

\bibitem{Adam2004}
Adam, J.: New correlations between electrical current and temperature rise in
  pcb traces.
\newblock In: Twentieth Annual IEEE Semiconductor Thermal Measurement and
  Management Symposium (IEEE Cat. No.04CH37545), pp. 292--299 (2004).
\newblock \doi{10.1109/STHERM.2004.1291337}

\bibitem{Adamatzky2014}
Adamatzky, A.: Towards plant wires.
\newblock Biosystems \textbf{122}, 1--6 (2014).
\newblock \doi{https://doi.org/10.1016/j.biosystems.2014.06.006}.
\newblock
  \urlprefix\url{http://www.sciencedirect.com/science/article/pii/S0303264714000872}

\bibitem{Adamatzky2017}
Adamatzky, A.: Physical Maze Solvers. All Twelve Prototypes Implement 1961 Lee
  Algorithm, pp. 489--504.
\newblock Springer International Publishing, Cham (2017).
\newblock \doi{10.1007/978-3-319-46376-6_23}.
\newblock \urlprefix\url{http://doi.org/10.1007/978-3-319-46376-6_23}

\bibitem{Assis2007}
Assis, A., Hernandes, J.: The Electric Force of a Current: {W}eber and the
  Surface Charges of Resistive Conductors Carrying Steady Currents.
\newblock Apeiron (2007).
\newblock \urlprefix\url{http://books.google.fr/books?id=YQdPHAAACAAJ}.
\newblock This book is freely available at the following address :
  \url{http://www.ifi.unicamp.br/~assis/}.

\bibitem{Ayrinhac2014}
Ayrinhac, S.: Electric current solves mazes.
\newblock Physics Education \textbf{49}(4), 443 (2014).
\newblock \urlprefix\url{http://stacks.iop.org/0031-9120/49/i=4/a=443}

\bibitem{Besson2009}
Besson, U.: Paradoxes of thermal radiation.
\newblock European Journal of Physics \textbf{30}(5), 995 (2009).
\newblock \urlprefix\url{http://stacks.iop.org/0143-0807/30/i=5/a=008}

\bibitem{Chabaybook}
Chabay, R.W., Sherwood, B.A.: Matter and interactions II: Electric and magnetic
  interactions.
\newblock Wiley, New York (2002)

\bibitem{Fuerstman2003}
Fuerstman, M.J., Deschatelets, P., Kane, R., Schwartz, A., Kenis, P.J.A.,
  Deutch, J.M., Whitesides, G.M.: Solving mazes using microfluidic networks.
\newblock Langmuir \textbf{19}(11), 4714--4722 (2003).
\newblock \doi{10.1021/la030054x}.
\newblock \urlprefix\url{https://doi.org/10.1021/la030054x}

\bibitem{Gross2005}
Gross, N.A., Hersek, M., Bansil, A.: Visualizing infrared phenomena with a
  webcam.
\newblock American Journal of Physics \textbf{73}(10), 986--990 (2005).
\newblock \doi{10.1119/1.1900105}.
\newblock \urlprefix\url{http://doi.org/10.1119/1.1900105}

\bibitem{Haglund2016}
Haglund, J., Jeppsson, F., Melander, E., Pendrill, A.M., Xie, C.,
  Sch\"{o}nborn, K.J.: Infrared cameras in science education.
\newblock Infrared Physics \& Technology \textbf{75}, 150 -- 152 (2016).
\newblock \doi{http://dx.doi.org/10.1016/j.infrared.2015.12.009}.
\newblock
  \urlprefix\url{http://www.sciencedirect.com/science/article/pii/S1350449515301985}

\bibitem{Heald1984}
Heald, M.A.: Electric fields and charges in elementary circuits.
\newblock American Journal of Physics \textbf{52}(6), 522--526 (1984).
\newblock \doi{10.1119/1.13611}.
\newblock \urlprefix\url{https://doi.org/10.1119/1.13611}

\bibitem{Jackson1996}
Jackson, J.D.: Surface charges on circuit wires and resistors play three roles.
\newblock American Journal of Physics \textbf{64}(7), 855--870 (1996).
\newblock \doi{10.1119/1.18112}.
\newblock \urlprefix\url{https://doi.org/10.1119/1.18112}

\bibitem{Liebmann1950}
Liebmann, G.: Solution of partial differential equations with a resistance
  network analogue.
\newblock British Journal of Applied Physics \textbf{1}(4), 92--103 (1950).
\newblock \urlprefix\url{http://stacks.iop.org/0508-3443/1/i=4/a=303}

\bibitem{Mollman2007}
M\"{o}llmann, K.P., Vollmer, M.: Infrared thermal imaging as a tool in
  university physics education.
\newblock European Journal of Physics \textbf{28}(3), S37--S50 (2007).
\newblock \urlprefix\url{http://stacks.iop.org/0143-0807/28/i=3/a=S04}

\bibitem{Muller2012}
M{\"u}ller, R.: A semiquantitative treatment of surface charges in dc circuits.
\newblock American Journal of Physics \textbf{80}(9), 782--788 (2012).
\newblock \doi{10.1119/1.4731722}.
\newblock \urlprefix\url{https://doi.org/10.1119/1.4731722}

\bibitem{Nakagaki2000}
Nakagaki, T., Yamada, H., T{\'o}th, {\'A}.: Intelligence: Maze-solving by an
  amoeboid organism.
\newblock Nature \textbf{407}(6803), 470--470 (2000).
\newblock \urlprefix\url{http://dx.doi.org/10.1038/35035159}

\bibitem{Netzell2017}
Netzell, E., Jeppsson, F., Haglund, J., Sch{\"o}nborn, K.J.: Visualising energy
  transformations in electric circuits with infrared cameras.
\newblock School Science Review \textbf{98}(364), 19--22 (2017)

\bibitem{Oh2012}
Oh, K.W., Lee, K., Ahn, B., Furlani, E.P.: Design of pressure-driven
  microfluidic networks using electric circuit analogy.
\newblock Lab Chip \textbf{12}, 515--545 (2012).
\newblock \doi{10.1039/C2LC20799K}.
\newblock \urlprefix\url{http://dx.doi.org/10.1039/C2LC20799K}

\bibitem{Pershin2011}
Pershin, Y.V., Di~Ventra, M.: Solving mazes with memristors: A massively
  parallel approach.
\newblock Phys. Rev. E \textbf{84}, 046,703 (2011).
\newblock \doi{10.1103/PhysRevE.84.046703}.
\newblock \urlprefix\url{http://link.aps.org/doi/10.1103/PhysRevE.84.046703}

\bibitem{Preyer2002}
Preyer, N.W.: Transient behavior of simple rc circuits.
\newblock American Journal of Physics \textbf{70}(12), 1187--1193 (2002).
\newblock \doi{10.1119/1.1508444}.
\newblock \urlprefix\url{https://doi.org/10.1119/1.1508444}

\bibitem{Qin2007}
Qin, J., Wheeler, A.R.: Maze exploration and learning in c. elegans.
\newblock Lab Chip \textbf{7}, 186--192 (2007).
\newblock \doi{10.1039/B613414A}.
\newblock \urlprefix\url{http://dx.doi.org/10.1039/B613414A}

\bibitem{Rainson1994}
Rainson, S., Transtr\"{o}mer, G., Viennot, L.: Students' understanding of
  superposition of electric fields.
\newblock American Journal of Physics \textbf{62}(11), 1026--1032 (1994).
\newblock \doi{10.1119/1.17701}.
\newblock \urlprefix\url{http://dx.doi.org/10.1119/1.17701}

\bibitem{Reyes2002}
Reyes, D.R., Ghanem, M.M., Whitesides, G.M., Manz, A.: Glow discharge in
  microfluidic chips for visible analog computing.
\newblock Lab Chip \textbf{2}, 113--116 (2002).
\newblock \doi{10.1039/B200589A}.
\newblock \urlprefix\url{http://dx.doi.org/10.1039/B200589A}

\bibitem{Rosser1970}
Rosser, W.G.V.: Magnitudes of surface charge distributions associated with
  electric current flow.
\newblock American Journal of Physics \textbf{38}(2), 265--266 (1970).
\newblock \doi{10.1119/1.1976298}.
\newblock \urlprefix\url{http://doi.org/10.1119/1.1976298}

\bibitem{Stasiek2014}
Stasiek, J., Jewartowski, M., Kowalewski, T.A.: The use of liquid crystal
  thermography in selected technical and medical applications -- recent
  development.
\newblock Journal of Crystallization Process and Technology \textbf{4}(1),
  46--59 (2014).
\newblock \doi{10.4236/jcpt.2014.41007}.
\newblock \urlprefix\url{http://dx.doi.org/10.4236/jcpt.2014.41007}

\bibitem{Stratton1973}
Stratton, L.O., Coleman, W.P.: Maze learning and orientation in the fire ant
  (solenopsis saevissima).
\newblock Journal of comparative and physiological psychology \textbf{83}(1), 7
  (1973)

\bibitem{Tarassenko1991}
Tarassenko, L., Blake, A.: Analogue computation of collision-free paths.
\newblock In: Proceedings. 1991 IEEE International Conference on Robotics and
  Automation, pp. 540--545 vol.1 (1991).
\newblock \doi{10.1109/ROBOT.1991.131636}

\bibitem{Vollmerbook}
Vollmer, M., M{\"o}llmann, K.: Infrared Thermal Imaging: Fundamentals, Research
  and Applications.
\newblock Wiley (2011).
\newblock \urlprefix\url{http://books.google.fr/books?id=b-MqbyPwAuoC}

\bibitem{Vollmer2001}
Vollmer, M., M{\"o}llmann, K.P., Pinno, F., Karst{\"a}dt, D.: There is more to
  see than eyes can detect.
\newblock The Physics Teacher \textbf{39}(6), 371--376 (2001).
\newblock \doi{http://dx.doi.org/10.1119/1.1407135}.
\newblock
  \urlprefix\url{http://scitation.aip.org/content/aapt/journal/tpt/39/6/10.1119/1.1407135}

\bibitem{Xie2011a}
Xie, C.: Visualizing chemistry with infrared imaging.
\newblock Journal of Chemical Education \textbf{88}(7), 881--885 (2011).
\newblock \doi{10.1021/ed1009656}.
\newblock \urlprefix\url{http://dx.doi.org/10.1021/ed1009656}

\bibitem{Xie2011}
Xie, C., Hazzard, E.: Infrared imaging for inquiry-based learning.
\newblock The Physics Teacher \textbf{49}(6), 368--372 (2011).
\newblock \doi{10.1119/1.3628268}.
\newblock \urlprefix\url{http://doi.org/10.1119/1.3628268}

\bibitem{Zhang2000}
Zhang, S., Mizutani, A., Srinivasan, M.V.: Maze navigation by honeybees:
  Learning path regularity.
\newblock Learning \& Memory \textbf{7}(6), 363--374 (2000).
\newblock \doi{10.1101/lm.32900}.
\newblock \urlprefix\url{http://learnmem.cshlp.org/content/7/6/363.abstract}

\end{thebibliography}
\end{document}